\documentstyle[amstex,12pt]{article}
\newtheorem{th}{\large\sc Theorem}[section]
\newtheorem{pro}{\large\sc Proposition}[section]
\newtheorem{defi}{\large\sc Definition}[section]

\begin{document}
\title{\sc AKNS Hierarchy, Self--Similarity, String Equations and the
Grassmannian\thanks{Partially supported by CICYT proyecto PB89-0133}}
\author{Francisco Guil\\ Departamento de F\'{\i}sica Te\'orica\\ Universidad
Complutense\\ E28040--Madrid\\ Spain\\and\\
Manuel Ma\~nas\thanks{Research supported by British
Council's Fleming award ---postdoctoral MEC   fellowship GB92 00411668, and
postdoctoral
EC Human Capital and Mobility individual fellowship ERB40001GT922134}\\  The
Mathematical Institute\\
Oxford University\\ 24-29 St.Giles', Oxford OX1 3LB \\United Kingdom}
\maketitle
\begin{abstract}
In this paper the Galilean, scaling and translational self--similarity
conditions for the AKNS hierarchy are analysed geometrically in terms of the
infinite dimensional Grassmannian. The string equations found recently by
non--scaling limit analysis of the one--matrix model are shown to correspond to
the Galilean self--similarity condition for this hierarchy.
We describe, in terms of the initial data for the zero--curvature 1--form of
the AKNS hierarchy, the moduli space of these self--similar solutions in the
Sato Grassmannian. As a byproduct we characterize the points in the
Segal--Wilson Grassmannian corresponding to the Sachs rational solutions of the
AKNS equation and  to the Nakamura--Hirota rational solutions of the NLS
equation. An explicit 1--parameter family of Galilean self--similar solutions
of the AKNS equation and the associated solution to the NLS equation is
determined.
\end{abstract}

\section{Introduction}

Matrix models have been extensively   used  as a non--perturbative formulation
of string theory. In the Hermitian one--matrix model with even potentials
\cite{her},  the double scaling limit implies for the specific heat   the
Korteweg--de Vries hierarchy and an additional constraint, the so called string
equation. This is relevant in the Witten--Kontsevich \cite{wk} description of
the the intersection theory of the moduli space of complex curves. Motivated by
some anomalous behaviour of the solutions to the string equation, a
modification of it was proposed in \cite{sta}, the 2D--stable quantum gravity.
The former string equation corresponds to  invariance under Galilean
transformations and the later to invariance under scaling transformations.
Further, it was shown \cite{un} that for the symmetric unitary matrix model
with even potentials and some boundary terms in the double scaling limit the
specific heat satisfies the modified Korteweg--de Vries hierarchy and a string
equation, corresponding to the self--similarity condition under
scaling   transformations.

The infinite--dimensional Grassmannian model of Sato \cite{sa} for the
Korteweg--de Vries hierarchy and the associated periodic flag manifold have
been extensively used in the analysis   of these string equations,
\cite{ks,gm2}.

The interplay of  matrix models with different integrable systems is of great
interest.
Very recently   a non--scaling limit analysis of the Hermitian one--matrix
model not restricted to the even potential case has been given, \cite{bx}. It
is found that the Ablowitz--Kaup--Newell--Segur (AKNS) hierarchy appears
naturally in the model besides a string equation. The Korteweg--de Vries
hierarchy is contained in the AKNS hierarchy as a reduction, and therefore
appears in the model before one takes the double scaling limit. The AKNS
hierarchy is a complexified version of the Nonlinear Schr\"odinger (NLS)
hierarchy and also contains the modified Korteweg--de Vries hierarchy. In
\cite{bx} one can find a discussion of the topological field theory associated
with the AKNS hierarchy.
The mentioned string equation   corresponds, as we shall show, to a Galilean
self--similarity condition for the AKNS hierarchy.

Because the AKNS hierarchy is not so well--known as the Korteweg--de Vries
and that we shall deal with it  along this paper, let us present now some facts
about it. In \cite{akns}   this hierarchy was first used implicitly to solve a
number of equations by a multicomponent inverse scattering method or inverse
spectral transform \cite{ac}. But the hierarchy appeared explicitly in
\cite{fnr} where it was extensively studied  \cite{ft,n}. In \cite{du} the
finite gap solutions were analysed  and for the real version, the NLS
hierarchy, this was done  in \cite{pre}. One can express these solutions in
terms of theta functions for the corresponding hiperelliptic curve. In the
papers \cite{d,w} a detailed account of the Grassmannian model, Baker and
$\tau$--functions can be found. In addition, in \cite{bk} the Toda--AKNS
hierarchy and its $\tau$--functions were introduced from the point of view of
representation theory for affine Lie algebras and the Birkhoff factorization
method!
. Notice also the similarity with
the hierarchy appearing in \cite{bx}.

 In this paper we analyse the  Sato Grassmannian geometry of the moduli space
of solutions to the string equation of the non--scaling limit of the
one--matrix model, and more generally of self--similar solutions under any of
the local symmetries of the AKNS hierarchy. These are Galilean, scaling and
translational transformations.
  We give a parametrization of this moduli space in terms of the initial
condition for the zero--curvature 1--form of the AKNS hierarchy. As a byproduct
we obtain the points in the Segal--Wilson Grassmannian corresponding to the
weighted scaling self--similar rational solutions of \cite{s,nh},  and we find
a 1--parameter family of Galilean self--similar solutions to the AKNS
hierarchy.

In the second section we introduce the AKNS and $\text{NLS}^{\pm}$
hierarchies.   We prove that a Galilean self--similar solution is self--similar
under certain weighted scaling, with the scaling weights determined by the
initial data for some associated conserved densities. We   present also a
zero--curvature type formulation of the string equations.

In the following section we consider the Birkhoff factorization problem for the
AKNS hierarchy and its relation with the Grassmannian. There we formulate the
two main results of the paper. The first one determines the stucture of the
initial conditions for which the Birkhoff factorization problem implies
self--similar solutions, and the second giving the structure of the set of
points in the Grassmannian associated with solutions to the string equations.
That is, we analyse the moduli space in the Grassmannian.

Finally, in section 4 we examine several examples. We consider the mixed
Galilean and translational self--similar condition, which corresponds to
Galilean self--similarity in appropriate shifted coordinates. We obtain points
that do not belong to the Segal--Wilson Grassmannian but to the Sato
Grassmannian and can be expressed in terms of Gaussian and Weber's parabolic
cylinder functions. We also give a family of Galilean self--similar solutions
of the AKNS equation and the corresponding reduction to the $\text{NLS}^+$
equation, an explode--decay non--localized wave. The scaling case with
different weights is also considered  in shifted coordinates. Now, there are
some points that belong to the Segal--Wilson Grassmannian, they correspond to
the rational solutions of \cite{s} for the AKNS equation, and some of them
reduces to the $\text{NLS}^+$ equation giving the rational solutions of
\cite{nh}. The subspaces in the Sato Grassmannian can be expressed in terms of
Tricomi--Kumm!
er's hipergeometric confluent func
tions that, in the mentioned rational case, are Laurent polynomials.

   \section{AKNS hierarchy and string
equations}\setcounter{equation}{0}\setcounter{th}{0}
\setcounter{pro}{0}

We begin this section with the definition of the integrable equations known as
the  AKNS hierarchy, which is a complexified version of the   NLS hierarchy. It
is defined
in terms of a couple of scalar functions $p,q$  that depend on an infinite
number of
variables ${\bf t}:=\{t_n\}_{n\geq 0}\in\Bbb{C}^{\infty}$ which are   local
coordinates for
the time manifold $\cal T$. In this convention we adopted $t_1$ to be the space
coordinate, usually denoted by $x$,  and $t_n$ with $n>1$ corresponds to a time
variable. The coordinate $t_0$  as we shall see  below is associated to a
symmetry of the standard AKNS hierarchy ($n>1$).

\begin{defi}
The AKNS  hierarchy for $p,q$ is the following collection of
compatible equations
\[
\begin{cases}
\partial_np= 2p_{n+1},\\
\partial_nq=-2q_{n+1},
\end{cases}
\]
where $ n\geq 0$,
 $\partial_n:=\partial/\partial t_n$
and $p_n,q_n$ and $h_n$ are defined recursively by the relations
\begin{eqnarray*}
&&p_n=\frac{1}{2}\partial_1p_{n-1}+ph_{n-1},\\
&&q_n=-\frac{1}{2}\partial_1q_{n-1}+qh_{n-1},\\
&&\partial_1h_n=pq_n-qp_n,\;\; n\geq 1
\end{eqnarray*}
with the initial data
\[
 p_0=q_0=0,\, h_0=1.
\]
  \end{defi}
{}From the recurrence relations one gets for example
\begin{eqnarray*}
&&p_1=p,\, q_1=q,\, h_1=0\\
&&p_2=\frac{1}{2}\partial_1p,\, q_2=-\frac{1}{2}\partial_1q,\,
h_2=-\frac{1}{2}pq\\
&&p_3=\frac{1}{4}\partial_1^2p-\frac{1}{2}p^2q,\,
q_3=\frac{1}{4}\partial_1^2q-\frac{1}{2}pq^2,\,
h_3=\frac{1}{4}(p\partial_1q-q\partial_1p).
\end{eqnarray*}

The $n=0$ flow is   usually not considered in the standard AKNS hierarchy, but
its inclusion will prove   quite convenient.  The equations for that flow are
\[
\begin{cases}
\partial_0p=2p,\\
\partial_0q=-2q,
\end{cases}
\]
which means that
\[
p(t_0,t_1,\dots)=\exp(2t_0)\tilde p(t_1,\dots),\;\,
q(t_0,t_1,\dots)=\exp(-2t_0)\tilde q(t_1,\dots).
\]
 The functions $(\tilde p,\tilde q)$ satisfy the standard AKNS hierarchy, and
this $t_0$--flow reflects the fact that given a solution  $(\tilde p,\tilde q)$
to the standard AKNS hierarchy ($n>0$) then  $(e^c\tilde p,e^{-c}\tilde q)$ is
a solution as well for any $c\in{\Bbb C}$.
The $n=1$ flow is an identity.

For $n=2$ the equations are
\[
\begin{cases}
2\partial_2p=\partial_1^2p-2p^2q,\\
2\partial_2q=-\partial_1^2q+2pq^2,
\end{cases}
\]
and for $n=3$ one has
\[
\begin{cases}
4\partial_3p=\partial_1^3p-6pq\partial_1p,\\
4\partial_3q=\partial_1^3q-6pq\partial_1q.
\end{cases}
\]
The principal reduction $p=q=v$ implies the modified Korteweg--de Vries
equation
$4\partial_3v=\partial_1^3v-6v^2\partial_1v$, and the reduction defined by
$p=1$ and $q=-u$ determines the Korteweg--de Vries equation
$\partial_3u=\partial_1^3u+6u\partial_1u$. Observe also that when $p=0$  one
obtains the heat equation $2\partial_2p=\partial_1^2p$, which is a particular
case of the heat hierarchy $2^{n-1}\partial_np=\partial_1^np$.

{}From the recurrence relations one easily deduces that
\[
\partial_1h_{n+1}=\partial_nh_2,
\]
from where it follows that $h_{n+1}=\partial_n\partial_1^{-1}h_2$ and so
\begin{equation}
\partial_mh_{n+1}=\partial_nh_{m+1},\label{cons}
\end{equation}
giving an infinity set of non--trivial local conservation laws \cite{d2}.

Notice that the real reduction $q=\mp p^{\ast}$ and $t_n\mapsto it_n$ produces
the $\text{NLS}^{\pm}$ hierarchy for which the $t_2$--flow is
$2i\partial_2 p=-\partial_1^2p\pm 2|p|^2p$, the $\text{NLS}^{\pm}$ equation,
and the $t_3$--flow is $4\partial_3p=-\partial_1^3p\pm 6|p|^2\partial_1p$.

\begin{defi}
The $\text{NLS}^{\pm}$ hierarchy
\[
i\partial_np=2p_{n+1}
\]
is defined in terms of the recursion relations
\begin{eqnarray*}
&&p_n=\frac{i}{2}\partial_1p_{n-1}+ph_{n-1}\\
&&\partial_1h_n=\mp2{\rm Im}\, pp_n^{\ast},
\end{eqnarray*}
and $p_0=0,h_0=1$.
\end{defi}

An essential feature of the AKNS hierarchy relies in its zero--curvature
formulation \cite{akns,fnr,n}. If $\{E,H,F\}$ is the standard Weyl--Cartan
basis for the simple Lie algebra ${\frak sl}(2,{\Bbb C})$ of $2\times 2$
complex, traceless matrices we define
\[
Q_n:=p_nE+h_nH+q_nF,
\]
 and denote by
\[
L_n(\lambda):=\sum_{m=0}^n\lambda^m Q_{n-m},
\]
where $\lambda$ is a complex spectral parameter. Introducing the differential
1--form
\[
\chi=\sum_{n\geq 0} L_ndt_n,
\]
one is allowed to formulate
  the AKNS hierarchy as the zero--curvature condition
\[
[d-\chi,d-\chi]=0,
\]
where $d$ is the exterior derivative operator on the differential forms
$\Lambda{\cal T}$.
This aspect of the AKNS hierarchy is connected with the   spectral problem
\[
\partial_1\Psi=\left(\begin{array}{cc}\lambda&p\\q&-\lambda\end{array}\right)\Psi,
\]
where
\[
\Psi=\left(\begin{array}{c}\psi_1\\ \psi_2\end{array}\right).
\]

For the $\text{NLS}^{\pm}$ hierarchy one has also a zero--curvature
formulation. Now the $Q_n=p_nE+ih_nH\mp p_n^{\ast}F$ are maps from ${\cal T}$
into the real Lie algebras
${\frak su}(2)$ and ${\frak su}(1,1)$ respectively.

 Let us now describe the local symmetries of the hierarchy. First we have the
shifts in the time variables, the infinite set of
translational symmetries are   isospectral
symmetries of the hierarchy in the sense that they preserve the
associated spectral problem.
  Let $\vartheta$ be
\[
\vartheta({\bf t}):={\bf t}+\mbox{\boldmath$\theta$},
\]
the action of translations, where
\[
\mbox{\boldmath$\theta$}:=\{\theta_n\}_{n\geq 0}
\in{\Bbb C}^{\infty},
\]
are the shifts of the time variables.

We have a local
action of the abelian group ${\Bbb C}^{\infty}$ on the time manifold
$\cal T$, then it follows
\begin{pro}
If $(p,q)$ is a solution
to the AKNS hierarchy  then so is $(\vartheta^{\ast}p,\vartheta^{\ast}q)$.
\end{pro}
 But there are also two local non--isospectral symmetries. One is the scaling
symmetry, and the other is the Galilean symmetry. Next we define both of them
 \begin{defi}
The Galilean  transformation  ${\bf t}\mapsto\gamma_a({\bf t})$ is given by
\[
\gamma_a({\bf t})_n:=
\sum_{m\geq 0}
\binom{n+m}m
a^mt_{n+m}
\]
where $a\in{\Bbb C}$.

The scaling transformation ${\bf t}\mapsto\varsigma_b({\bf t})$ is represented
by the relations
\[
\varsigma_b({\bf t})_n:=e^{nb}t_n
\]
where $b\in{\Bbb C}$.
\end{defi}

We have  two additive local actions of $\Bbb C$ over
$\cal T$. One can   show that
\begin{pro}
If $(p,q)$ is a
 solution of the AKNS hierarchy  then so are
$(\gamma_a^{\ast}p,\gamma_a^{\ast}q)$ and
$(e^b\varsigma_b^{\ast} p,e^b\varsigma_b^{\ast}q)$.
\end{pro}
It proves  convenient to define
\[
t(\lambda):=\sum_{n\geq 0}\lambda^nt_n.
\]
Observe that for the isospectral symmetries we have
\[
\vartheta^{\ast}t(\lambda)=t(\lambda)+\theta(\lambda)
\]
where
\[
\theta(\lambda):=\sum_{n\geq 0}\theta_n\lambda^n,
\]
and that for the non--isospectral symmetries one has
\[
\gamma_a^{\ast}t(\lambda)=t(\lambda+a),\,\;
\varsigma_b^{\ast}t(\lambda)=t(e^b\lambda).
\]

Notice that for the corresponding solutions $(\tilde p,\tilde q)$ of the
standard AKNS hierarchy  the Galilean action is
$(\exp(2t(a))\gamma_a^{\ast}\tilde p,(\exp(-2t(a))\gamma_a^{\ast}\tilde q)$,
the exponential factors are a result of the flow in $t_0$ induced by the
Galilean transformation.
The related fundamental vector fields, infinitesimal generators of the
action of translation, Galilean and scaling   transformations are
given by
\[
\partial_n,\;n\geq 0,
\,\;  \mbox{\boldmath$\gamma$}=\sum_{n\geq 0}(n+1)
t_{n+1}\partial_n,\,\;\mbox{\boldmath$\varsigma$}
=\sum_{n\geq 1}n t_n\partial_n,
\]
respectively. They generate the linear space ${\Bbb C}\{ \partial_n,
\mbox{\boldmath$\varsigma$},\mbox{\boldmath$\gamma$} \}_{n\geq 0 }$ which  is
the Lie algebra of local symmetries of the AKNS hierarchy, the non--trivial Lie
brackets are
\[
[\partial_n,\mbox{\boldmath$\varsigma$}]=n \partial_n,\;\,
[\partial_{n+1},\mbox{\boldmath$\gamma$}]=(n+1)\partial_n,\,\;
 [\mbox{\boldmath$\varsigma$},\mbox{\boldmath$\gamma$}]=
2\boldsymbol{\gamma}.
\]

Consider the following vector field belonging to this
Lie algebra,
\[
X:=
\mbox{\boldmath$\vartheta$}+ a\mbox{\boldmath$\gamma$}
+b\mbox{\boldmath$\varsigma$},
\]
with
\[
\mbox{\boldmath$\vartheta$}=\sum_{n\geq 0} \theta_n
\partial_n,
\]
defining  a superposition of translations,
Galilean and scaling
transformations.

If $(p,q)$ is a solution of the AKNS hierarchy then there is a 1--parameter
family of solutions $(p_{\tau},q_{\tau})$   generated by the vector field $X$.
We have the  important notion
\begin{defi}
 A self--similar solution under any of the mentioned symmetries is a solution
which remains invariant under the corresponding transformation.
\end{defi}
  Then we have,
\begin{pro}
A solution $(p,q)$ of the AKNS hierarchy
is self--similar under  the action of the vector field $X$
if and only if
it satisfies the generalized string equations
\begin{equation}
\begin{cases}
Xp+bp=0, \\
Xq+bq=0,
\end{cases}
 \label{sssecu}
\end{equation}
\end{pro}
 Notice that when $X=\boldsymbol{\gamma}+\sum_{n\geq 0}\theta_n\partial_n$ one
can perform the coordinate transformation
$t_{n+1}\mapsto t_{n+1}+\theta_n/(n+1)$. Thus, the coefficient $\theta_n$ is
equivalent to a shift in the time coordinate $t_{n+1}$.

Now, if $X=\boldsymbol{\varsigma}+\sum_{n\geq 0}\theta_n\partial_n$ we can
define the transformation $t_{n+1}\mapsto t_{n+1}+\theta_{n+1}/(b(n+1))$ and
obtain in the new coordinates a
vector field corresponding to scaling and a term of type $\theta_0\partial_0$.
This last term can be understood as follows.
Given a solution $(p,q)$ to the  AKNS hierarchy
then $(\exp(b(1+2\theta_0))\varsigma_b^{\ast}p,
(\exp(b(1-2\theta_0))\varsigma_b^{\ast}q)$ is a solution as well. So solutions
self--similar under the vector field $X$ correspond in adequate coordinates, to
self--similarity under this particular scaling, that we shall call
$(1+2\theta_0,1-2\theta_0)$ weighted scaling.

Now we shall prove that Galilean self--similarity implies scaling
self--similarity. We have,

\begin{pro}
If $(p,q)$ is a    solution to the AKNS hierarchy self--similar under the
action of the vector field
 \[
\boldsymbol{\gamma}+\sum_{n\geq 0}\theta_n\partial_n,
\]
then it is also self--similar under the action of the vector field
\[
\boldsymbol{\varsigma}+\sum_{n\geq 0}\theta_n \partial_{n+1}-(\sum_{n\geq
1}\theta_n h_{n+1}\left|_{{\bf t}=0}\right.)\partial_0.
\]
\end{pro}
This proposition simply says that the $L_{-1}$--Virasoro constraint implies the
$L_0$--Virasoro constraint.

{\bf Proof:}
We have
\[
(\boldsymbol{\gamma}+\sum_{n\geq
0}\theta_n\partial_n)p=(\boldsymbol{\gamma}+\sum_{n\geq
0}\theta_n\partial_n)q=0.
\]
Therefore, we obtain the relations
\[
(\boldsymbol{\gamma}+\sum_{n\geq
0}\theta_n\partial_n)p_{n+1}=-np_n,\,\;(\boldsymbol{\gamma}+
\sum_{n\geq 0}\theta_n\partial_n)q_{n+1}=-nq_n,
\]
where, for example, we have used the fact that $2p_{n+1}=\partial_np$,  $p$ is
killed by
$\boldsymbol{\gamma}+\sum_{n\geq 0}\theta_n\partial_n$ and the commutation
relation of this vector field and $\partial_n$.
One can equally deduce
\[
(\boldsymbol{\gamma}+\sum_{n\geq 0}\theta_n\partial_n)h_{n+1}=-nh_n.
\]

Because
\[
\begin{cases}
\partial_{n+1}p=(\frac{1}{2}\partial_n+2h_{n+1})p\\
\partial_{n+1}q=-(\frac{1}{2}\partial_n+2h_{n+1})q
\end{cases}
\]
it follows
\[
\begin{cases}
(\boldsymbol{\varsigma}+\sum_{n\geq
0}\theta_n\partial_{n+1})p=\frac{1}{2}(\boldsymbol{\gamma}+\sum_{n\geq
0}\theta_n\partial_n)p+2(\sum_{n\geq 1}(nt_n+\theta_{n-1}) h_n)p\\
(\boldsymbol{\varsigma}+\sum_{n\geq
0}\theta_n\partial_{n+1})q=-\frac{1}{2}(\boldsymbol{\gamma}+
\sum_{n\geq 0}\theta_n\partial_n)q-2(\sum_{n\geq 1}(nt_n+\theta_{n-1}) h_n)q.
\end{cases}
\]
 Observe that
\[
\partial_n \sum_{m\geq 1}(mt_m+\theta_{m-1})
h_m=nh_n+(\boldsymbol{\gamma}+\sum_{m\geq 0}\theta_m\partial_m)h_{n+1},
\]
as follows from (\ref{cons}). Hence, when $(p,q)$ is self--similar under
$\boldsymbol{\gamma}+\sum_{m\geq 0}\theta_m\partial_m$ we have
\[
\sum_{n\geq 1}(nt_n+\theta_{n-1}) h_n=  \sum_{n\geq 0}
\theta_{n}h_{n+1}\left|_{{\bf t}=0}\right..
\]
  This implies
\[
\begin{cases}
(\boldsymbol{\varsigma}+\sum_{n\geq 0}\theta_n\partial_{n+1})p-2 (\sum_{n\geq
0} \theta_{n} h_{n+1}\left|_{{\bf t}=0}\right.)p=0\\
(\boldsymbol{\varsigma}+\sum_{n\geq 0}\theta_n\partial_{n+1})q+2(\sum_{n\geq 0}
\theta_{n} h_{n+1}\left|_{{\bf t}=0}\right.)q=0,
\end{cases}
\]
and the proposition follows.$\Box$

If we denote by $p=\exp(s)$ and $q=-u\exp(-s)$ then the AKNS hierarchy
transforms in the hierarchy appearing in \cite{bx} for the fields $u=R$ and
$S=\partial_1s$, and the string equation is the one above with $t_0=0$ and
$b=\theta_n=0$. This hierarchy appears in that papers as a result of a
non--scaling limit analysis of the Hermitian one--matrix model. The first
conserved density of the AKNS hierarchy
is proportional to
the specific heat
\[
2h_2=-pq=\partial_1^2\ln Z.
\]

If $a=b=0$ one is led to the translational self--similar solutions of the AKNS
hierarchy, that is, the finite--gap solutions of the integrable equation in the
spirit of Novikov. The solutions of that type can be constructed in terms of
Riemann surfaces, in particular hiperelliptic curves, and the corresponding
$\tau$ and Baker functions can be expressed in terms of theta functions of such
curves (see \cite{du,d} for the AKNS equation and \cite{pre} for the NLS
equation).   The Galilean   self--similarity condition in the KdV case is
considered by Novikov \cite{no} as a quantized version of the finite gap
solutions.

 In  general the self--similarity condition can be reformulated
 as a zero--curvature type
condition.   We  define the outer derivative
\begin{equation}
\delta:=(a+b\lambda)\frac{d}{d\lambda},\label{der}
\end{equation}
and
 \begin{equation}
M:=\langle\chi,X\rangle,\label{m}
\end{equation}
Here $\langle\cdot,\cdot\rangle$  is the standard pairing between
1--forms and vector fields.
Then one has,

\begin{th}
  The zero--curvature type condition
\begin{equation}
[d-\chi,\delta-M]=0\label{zcss}
\end{equation}
is equivalent to the string equation (\ref{sssecu}).
 \end{th}

{\bf Proof:} This follows from the condition
 \[
\delta\chi=L_X\chi,
\]
where $L_X$ denotes the Lie derivative along the vector field $X$.
But
\[
L_X\chi=(i_Xd+di_X)\chi,
\]
and  recalling the zero--curvature condition for $\chi$, we obtain the
desired result.   $\Box$

This theorem plays a key r\^ ole for the analysis of the moduli space of the
string equation and it is associated with the isomonodromony method.

All  results regarding symmetries can be reduced to the $\text{NLS}^{\pm}$
hierarchy with $\theta_n=i\tilde\theta_n$ and $\tilde\theta_n,a,b\in\Bbb{R}$.

  \section{Grassmannians and  the moduli space for the string equations}
\setcounter{equation}{0} \setcounter{th}{0}\setcounter{pro}{0}
In this section we use the Grassmannian manifold  $\mbox{Gr}^{(2)}$ to describe
 the
AKNS flows, and to characterize
geometrically the string equations for
the self--similar solutions of the AKNS hierarchy. This manifold appears when
one considers  the Birkhoff factorization problem.

Recall that $\chi$ defines a 1--form with values in the loop algebra
$L{\frak sl}(2,{\Bbb C})$ of smooth maps from the circle
$S^1:=\{\lambda\in{\Bbb C}: |\lambda|=1\}$ to
the simple Lie algebra ${\frak sl}(2,{\Bbb C})$.
We define  an infinite set of commuting flows in the corresponding
loop group $LSL(2,{\Bbb C})$
\[
\psi({\bf t},\lambda):=\exp(t(\lambda)H)\cdot g(\lambda)
\]
where $g$ is the
initial condition.
  Denote by $L^+SL(2,{\Bbb C})$ those loops which have a holomorphic
extension to the interior of $S^1$ \cite{ps}, and by
$L^-_1SL(2,{\Bbb C})$ those which
extend analitically to the exterior of the circle and  are normalized
by the identity at $\infty$.

The Birkhoff factorization problem for a given $\psi({\bf t})$ consists in
finding the representation
\begin{equation}
\psi=\psi_-^{-1}\cdot\psi_+,\label{fac}
\end{equation}
where $\psi_-\in L^-_1SL(2,{\Bbb C})$ and $\psi_+\in
L^+SL(2,{\Bbb C})$,
and is  connected with the AKNS hierarchy.
The element $\psi_-$ can be parametrized by  functions $(p,q)$ in such a way
that $\psi_-$ is a solution to the
factorization problem if and only if $(p,q)$ is a solution to the  AKNS
hierarchy \cite{gm1}.
To this end one factorizes $\psi_-$ as follows
\[
\psi_-=u\cdot\phi
\]
where
\[
\ln u=\sum_{n\geq 1}\lambda^{-n}U_n,\,\;\phi=\exp(\sum_{n\geq
1}\Phi_n\lambda^{-n}H)
\]
 here $U_n({\bf t})\in\text{Im}\,\text{ad}H$ and $\partial_m\Phi_n$ can be
expressed as polynomials on $(p,q)$ and its $\partial_1$--derivatives.
For the elements of Sec. 2
we have the relation
\[
Q_n=\sum\Sb i_1+\cdots+i_m=n \\ 1\leq m\leq n\endSb\frac{1}{
m!}\text{ad}U_{i_1}\cdots\text{ad}U_{i_m}H,\]
and an infinite set of non--trivial local conservation laws given by
\[
\partial_n(\partial_1\Phi_m)=\partial_1(\partial_n\Phi_m),
\]
for the evolution generated by the vector field $\partial_n$. This conservation
laws where first found in \cite{ds} and  are equivalent to the $h_n$ of
\cite{d2} as is shown in \cite{bk}.

The $n=0$ flow is trivial, $\partial_0\Phi_n=0$ and
$(\partial_0-\text{ad}H)U_n=0$.

One also has that
\begin{equation}
\chi:=d\psi_+\cdot\psi_+^{-1}=
P_+\mbox{Ad}\psi_-\left(Hdt(\lambda)\right)\label{faco}
\end{equation}
is
the zero--curvature 1--form for the AKNS hierarchy \cite{gm1}.
Here $\mbox{id}=P_++P_-$ is the resolution of the identity related to the
spliting
\[
L{\frak sl}(2,{\Bbb C})=L^+{\frak sl}(2,{\Bbb C})\oplus
L^-_1{\frak sl}(2,{\Bbb C}).
\]
Observe that
\begin{equation}
\text{Ad}\psi_-H=\sum_{n\geq 0}\lambda^{-n}Q_n.\label{qh}
\end{equation}

One can conclude from these considerations that
the projection of the commuting flows
$\psi({\bf t})$ on the Grassmannian
manifold \cite{ps,sw}
\[
LSL(2,{\Bbb C})/
L^+SL(2,{\Bbb C})\cong \mbox{Gr}^{(2)},
\]
can be described   in terms of the AKNS hierarchy \cite{gm1,w}.

The element $g$ determines a point in the Grassmannian manifold
up to the gauge freedom $g\mapsto g\cdot h$, where $h\in
L^+SL(2,{\Bbb C})$.
A solution of  the AKNS hierarchy
 does  not change when $g(\lambda)\mapsto\exp(\beta(\lambda)
H)\cdot g(\lambda)$
if $\exp(\beta H)\in L^-_1SL(2,{\Bbb C})$.
We can say that the  moduli
space for the AKNS hierarchy contains the double coset space
\[
{\cal M}:=\Gamma_-\backslash LSL(2,{\Bbb C})/L^+SL(2,{\Bbb C}) \]
where $\Gamma_-$
is the abelian subgroup with Lie algebra
${\Bbb C}\{\lambda^nH \}_{n<0}$.

This  makes a connection with the Grassmannian description   for the AKNS
hierarchy given in \cite{d,w}.
The Baker function $w({\bf t})\in LSL(2,{\Bbb C})$ corresponds to
\[
w=\psi_-\cdot\exp(tH)=\psi_+\cdot g^{-1}.
\]
If we introduce the notation
\[
g=\left(\begin{array}{cc}\varphi_1&\tilde\varphi_1\\
\varphi_2&\tilde\varphi_2\end{array}\right),
\]
then we have the associated subspace
\[
W={\Bbb C}\left\{\lambda^n\left(\tilde\varphi_2,\;-\tilde\varphi_1\right),
\lambda^n\left(\varphi_2,\;-\varphi_1\right)\right\}_{n\geq 0},
\]
with $\lambda W\subset W$, in the Grassmannian $\text{Gr}^{(2)}$, \cite{ps,sw}.
The Baker function is the unique function with its rows taking its values in
$W$ such that
$P_+(w\cdot\exp(-tH))=1$. Obviously we have
\[
\partial_1w=L_1w
\]
and also
\[
\partial_nw=L_nw.
\]
The rows of the adjoint Baker function $w^{\ast}=(w^{-1})^t$ are maps into the
subspace \[
W^{\ast}={\Bbb C}\left\{\lambda^n\Phi, \lambda^n\tilde \Phi\right\}_{n\geq
0}\in\text{Gr}^{(2)},
\]
where
\[
\Phi:=  (\varphi_1,\;\;\varphi_2),
\,\;\tilde\Phi:=  (\tilde\varphi_1,\;\;\tilde\varphi_2).
\]
We shall adopt this subspace as a representative of the coset $g\cdot
L^+SL(2,{\Bbb C})$.

Let us now try to find for which initial conditions $g$
one gets
self--similar solutions, i.e. points in the Grassmannian  that are connected to
self--similar
solutions of the AKNS
hierarchy.

Recall that we have the derivation $\delta
\in  \mbox{Der}L^+{\frak sl}(2,{\Bbb C})$ defined in (\ref{der})
and the vector  $M({\bf t})\in L^+{\frak sl}(2,{\Bbb C})$
defined in (\ref{m}).
One has the
  \begin{th}$\,$
   If the initial condition $g$ satisfies the
equation
\begin{equation}
\delta g\cdot g^{-1}+{\rm Ad}g K=(\theta+f)  H,
\label{ecu}
\end{equation}
for  some $K\in L^+{\frak sl}(2,{\Bbb C})$ and some $f\in L^-_1{\Bbb C}$,
then
the corresponding solution to the  AKNS hierarchy satisfies the
  string equation (\ref{sssecu}).
 \end{th}

{\bf Proof:}
 For $\chi=d\psi_+\cdot\psi_+^{-1}$  we observe that
the equation (\ref{zcss})
holds if and only if
\begin{equation}
M=\delta\psi_+\cdot\psi_+^{-1}+\mbox{Ad}\psi_+K,\label{mm}
\end{equation}
for some $K\in L^+{\frak sl}(2,{\Bbb C})$. This, together with
the factorization problem (\ref{fac}), implies the relation
\[
M=\delta\psi_-\cdot\psi_-^{-1}+\mbox{Ad}\psi_-\left(
(a+b\lambda)\frac{dt}{d\lambda}H+
\mbox{Ad}\exp(tH)(\delta g\cdot g^{-1}+\mbox{Ad}g K)\right).
\]
Now, $M({\bf t})\in L^+{\frak sl}(2,{\Bbb C})$ and Eq.(\ref{ecu})
gives
\[
M=P_+\mbox{Ad}\psi_-\left((a+b\lambda)\frac{dt}{d\lambda}+
\theta H\right).
\]
Taking into account Eq.(\ref{faco}) we recover  (\ref{m}) and therefore the
 string equation is satisfied.
  $\Box$

Notice that     the function $f$ can be transformed into
\[
f(\lambda)\mapsto f(\lambda)+(a+\lambda b)\frac{d\beta}{d\lambda}(\lambda),
\]
where $\beta\in L^-_1{\Bbb C}$. If $b\neq 0$ then one transforms $f\mapsto 0$,
but when $b=0, a\neq 0$  one is only allowed to do $f\mapsto c\lambda^{-1}$,
finally if $a=b=0$ we can not remove $f$.

The Sato Grassmannian \cite{sa} contains much more self--similar solutions than
the Segal--Wilson one \cite{sw}. In fact, only  the finite gap solutions
---pure translational self--similarity--- and  the scaling self--similar
rational solutions of Sachs \cite{s} for the AKNS equation, and the
corresponding Nakamura--Hirota solutions for $\text{NLS}^+$  equation
\cite{nh}, are found in this Grassmannian. Therefore, we shall consider the
Sato Grassmannian $\mbox{Gr}^{(2)}$.
The statements above, which are rigorous in the
Segal--Wilson
case, can be extended to the Sato frame if the formal group $L^-_1SL(2,{\Bbb
C})$ is considered only
when acting by its adjoint action or by gauge transformations
in the   Lie algebra
${\frak sl}(2,{\Bbb C})[[\lambda^{-1},\lambda]$.
In this context  Eqs.(\ref{zcss},\ref{mm},\ref{ecu})  still hold.

Notice that for each equivalence class in $\cal M$ an element $g$
can be taken such that  $\ln g\in {\frak sl}(2,{\Bbb C})[[\lambda^{-1})$,
and that any element in the coset $g\cdot L^+SL(2,{\Bbb C})$ gives the same
point in the Grassmannian. One has the

 \begin{th}
The subspace
 \[
W^{\ast}={\Bbb C}\left\{\lambda^n\Phi,
\lambda^n \tilde\Phi\right\}_{n\geq 0},
\]
with $\Phi(\lambda),\tilde\Phi(\lambda)\in{\Bbb C}^2$,  corresponds to a
self--similar solution of the AKNS hierarchy under the action of the vector
field $X=a{\boldsymbol \gamma}+b{\boldsymbol\varsigma}+\sum_{n\geq
0}\theta_n\partial_n$,
 if $\Phi,\tilde\Phi$ have the asymptotic expansion
\begin{eqnarray*}
&& \Phi(\lambda)\sim \left(\begin{array}{cc}
1+  \varphi_{11} \lambda^{-1}+ \cdots,&
\varphi_{21}\lambda^{-1}+\varphi_{22}\lambda^{-2}+\cdots\end{array}
\right),\,\;\lambda\rightarrow\infty \\
&&\tilde\Phi(\lambda)\sim \left(\begin{array}{cc}
\tilde\varphi_{11} \lambda^{-1}+\tilde\varphi_{12}\lambda^{-2}+ \cdots, &
1+\tilde\varphi_{21}\lambda^{-1}+\cdots\end{array}
\right),\,\;
\lambda\rightarrow\infty,
\end{eqnarray*}
and satisfy
\begin{enumerate}
\item
When  $b\neq 0$ the ordinary differential equations
\begin{eqnarray*}
&&(a+b\lambda)\frac{d\Phi}{d\lambda}+(\sum_{n,m\geq 0}\lambda^n
\theta_{n+m}h_{m,0})\Phi+(\sum_{n,m\geq 0}\lambda^n
\theta_{n+m}q_{m,0})\tilde\Phi=\theta(\lambda)\Phi H,\\
&&(a+b\lambda)\frac{d\tilde\Phi}{d\lambda}-(\sum_{n,m\geq 0}\lambda^n
\theta_{n+m}h_{m,0})\tilde\Phi+(\sum_{n,m\geq 0}\lambda^n
\theta_{n+m}p_{m,0})\Phi=\theta(\lambda)\tilde\Phi H.
\end{eqnarray*}
\item
When $b=0, a\neq 0$ the ordinary differential equations
\begin{eqnarray*}
&&a\frac{d\Phi}{d\lambda}+(\sum_{n,m\geq 0}\lambda^n
\theta_{n+m}h_{m,0})\Phi+(\sum_{n,m\geq 0}\lambda^n
\theta_{n+m}q_{m,0})\tilde\Phi=(\theta(\lambda)-\lambda^{-1} \sum_{n\geq
0}\theta_nh_{n+1,0})\Phi H,\\
&&a\frac{d\tilde\Phi}{d\lambda}-(\sum_{n,m\geq 0}\lambda^n
\theta_{n+m}h_{m,0})\tilde\Phi+(\sum_{n,m\geq 0}\lambda^n
\theta_{n+m}p_{m,0})\Phi=(\theta(\lambda)-\lambda^{-1} \sum_{n\geq
0}\theta_nh_{n+1,0})\tilde\Phi H,
\end{eqnarray*}
\item
And when $a=b=0$ the algebraic relations
\begin{eqnarray*}
&&(\sum_{n,m\geq 0}\lambda^n \theta_{n+m}h_{m,0})\Phi+(\sum_{n,m\geq
0}\lambda^n
\theta_{n+m}q_{m,0})\tilde\Phi=(\theta(\lambda)+f(\lambda))\Phi H,\\
&&-(\sum_{n,m\geq 0}\lambda^n \theta_{n+m}h_{m,0})\tilde\Phi+(\sum_{n,m\geq
0}\lambda^n
\theta_{n+m}p_{m,0})\Phi=(\theta(\lambda)+f(\lambda))\tilde\Phi H,
\end{eqnarray*}
where
\begin{equation}
f(\lambda)=\sqrt{-{\rm det}(\sum_{n\geq 0}\theta_n
L_{n,0}(\lambda))}-\theta(\lambda)=\sqrt{-{\rm det}(\sum\begin{Sb}n
>0\\m\geq 0\end{Sb}\theta_mQ_{n+m,0}\lambda^{-n})},\label{f}
\end{equation}
has the asymptotic expansion
\[
f(\lambda)\sim\sum _{n>0}f_n\lambda^{-n},\;\, \lambda\rightarrow\infty
\]
with the recursion relation
\[
f_n=-\sum_{m=1}^{n-2}h_{n-m,0}f_{m}-\sum_{m\geq 0}\theta_m h_{n+m,0}.
\]
\end{enumerate}

\end{th}
Here we denote $F_0=F\left|_{{\bf t}=0}\right.$.

 {\bf Proof:}
Since $\exp(tH)\left|_{{\bf t}=0}\right.=\mbox{id}$ it follows from
(\ref{fac}) that $\psi_+\left|_{{\bf t}=0}\right.=\mbox{id}$ (formally
$g^{-1}=\psi_-\left|_{{\bf t}=0}\right.$)
and Eq.(\ref{mm}) gives
\[
K=M\left|_{{\bf t}=0}\right..
\]
But, from (\ref{m}) we have
\[
K=\langle\chi\left|_{{\bf t}=0}\right.,\mbox{\boldmath$\vartheta$}\rangle,
\]
where we have taken into account that
\[
X\left|_{{\bf t}=0}\right.=\mbox{\boldmath$\vartheta$}.
\]
Observe that
\begin{equation}
K=\sum_{n\geq 0}\theta_nL_n\left|_{{\bf t}=0}\right.=\text{Ad}g^{-1}(\theta H)-
P_-\text{Ad}g^{-1}\theta H,\label{kk1}
\end{equation}
where we have used $(\psi_-\left|_{{\bf t}=0})^{-1}\right.=g$. Therefore, we
have
\begin{equation}
\text{Ad}gK=\theta H-\text{Ad}gP_-\text{Ad}g^{-1}\theta H,\label{kk}
\end{equation}

When $b\neq 0$  we can remove the function $f$, and from (\ref{ecu}) one gets
the desired result. When $b=0, a\neq 0$ we have a contribution from $f$ of type
$c\lambda^{-1}$. This can be handled as follows. With the aid of Eq.(\ref{kk})
the equation (\ref{ecu}) can be written as
\[
a\frac{dg}{d\lambda}\cdot g^{-1}-\text{Ad}gP_-\text{Ad}g^{-1}\theta
H=c\lambda^{-1}H.
\]
Now, because the residue at $\lambda=0$ of the first term on the left hand side
of the equation above vanishes we have
\[
-\text{res}_{\lambda=0}\text{Ad}g^{-1}\theta H=cH,
\]
or
\[
-\sum_{n\geq 0}\theta_nQ_{n+1,0}=cH,
\]
thus
\[
c=-\sum_{n\geq 0}\theta_nh_{n+1,0}.
\]

When $a=b=0$ the Eqs.(\ref{ecu},\ref{kk1} ) implies the form of $f$ in the
first equality of (\ref{f}), the second expression follows from
(\ref{kk},\ref{qh}).
With this the proof is completed.$\Box$

This theorem provides us with a parametrization of the moduli space of
self--similar solutions of the AKNS hierarchy under the action of the vector
field $X$ in terms of initial conditions for the zero--curvature 1--form
$\chi$.
Notice that the equation characterizing $g$ depends on $K=\sum_{n\geq
0}\theta_nL_n\left|_{{\bf t}=0}\right.$. Thus, if $\theta$ is a polynomial of
degree $N$ the matrix $K$ depends  on $3N$ constants $\{p_n,q_n,h_n\}_{n=1}^N$,
but the $h_n$ can be expressed as polynomials of $\{p_m,q_m\}_{m=1}^{n-1}$.
When $a$ or $b$ do not vanish  we have an inclusion of this $2N$--dimensional
algebraic variety into the Sato Grassmannian, but one of the parameters can be
supressed because the freedom $(p,q)\mapsto(e^cp,e^{-c}q)$. Thus, there is an
inclusion of a $2N-1$--dimensional algebraic variety into the Sato Grassmannian
providing us with a description of the moduli space. When $a=b=0$ one has the
additional dependence on $f$ which is a function of $K$ only, and therefore one
has an inclusion of that algebraic variety into the Segal--Wilson Grassmannian,
the finite--gap solutions associated with hiperelliptic curves.

\section{Examples}
We give in this section a concrete analysis of the ODE's characterizing the
points in the Grassmannian associated with self--similar solutions. We start
with the Galilean case and then we study the weighted scaling case. For the
Galilean case we see that the points corresponding to self--similar solutions
can be expressed in terms of Gaussian and Weber's parabolic cylinder functions,
and that they never belong to the Segal--Wilson Grassmannian but to the Sato
Grassmannian. We give the analytic expression of the solution for the AKNS
system when $t_n=0$ for $n>2$. In  the weighted scaling case we find that the
points in the Grassmannian can be constructed with the aid of Tricomi--Kummer's
confluent hipergeometric functions. We see that for certain cases, when the
rows of $g$ are Laurent polynomials of different degrees and therefore define
points in the Segal--Wilson Grassmannian, these points are associated to the
rational solutions of the AKNS equation found in \cite{s} and!
 to the corresponding rational sol
utions of the $\text{NLS}^+$ equation of \cite{nh}.

\subsection{Galilean self--similarity}

We are going  to consider the string equation defined by  the vector field
$X=\boldsymbol{\gamma}+\theta_1\partial_1$.
As we have already discussed this corresponds  to
self--similar solutions under the Galilean symmetry in the shifted coordinates
$t_2\mapsto t_2+\theta_1/2$ and $t_n\mapsto t_n$ for $n\neq 2$. This  shift
allows us to avoid the singularities of the solution at $t_2=0$.

The form of the initial condition is
\[
g=\text{id}+\lambda^{-1}X_1+\cdots ,
\]
  which corresponds  to a self--similar solution under the vector field $X$ if
it   satisfies
\begin{equation}
\frac{dg}{d\lambda}+g \theta_1(p_0 E+\lambda H+q_0F)  =
\theta_1\left(\lambda+\frac{p_0q_0}{2}\lambda^{-1}\right)Hg\label{ecugg}
\end{equation}
that  for   $X_n$ reads
\[
-\left(n+\theta_1\frac{p_0q_0}{2}H\right)X_n +\theta_1
X_{n+1}(p_0E+q_0F)=\theta_1[H,X_{n+2}].
\]
If we introduce the notation
\begin{equation}
X_n=\left(\begin{array}{cc} A_n&X_n^+\\X_n^-&B_n\end{array}\right)\label{not1}
\end{equation}
it results
\begin{eqnarray*}
&&X_{n+2}^+=-\frac{1}{2\theta_1}
\frac{\left(n+\frac{\theta_1}{2}p_0q_0\right)
\left(n+1+\frac{\theta_1}{2}p_0q_0\right)}{n+1}X_n^+\\
&&X_{n+2}^-=\frac{1}{2\theta_1}
\frac{\left(n-\frac{\theta_1}{2}p_0q_0\right)
\left(n+1-\frac{\theta_1}{2}p_0q_0\right)}{n+1}X_n^-
 \end{eqnarray*}
and
\begin{eqnarray*}
&&A_n=\frac{\theta_1q_0}{n+\frac{\theta_1}{2}p_0q_0}X_n^+\\
&&B_n=\frac{\theta_1p_0}{n-\frac{\theta_1}{2}p_0q_0}X_n^-
\end{eqnarray*}
that together with   $X_1^+=p_0/2$ and $X_1^-=-q_0/2$ gives us the matrix $g$.
Observe that $X_{2n}^+=X^-_{2n}=0$ and
$A_{2n+1}=B_{2n+1}=0$. The expansion never converges, we can choose $p_0q_0$
such that the first row of $g$ is polynomial in
$\lambda^{-1}$ but the the second row does not converge. We conclude that this
solution belongs to the Sato Grassmannian and not to the Segal--Wilson one.

Now, writing
\begin{equation}
g=\left(\begin{array}{cc}A&X^+\\X^-&B\end{array}\right)\label{not2}
\end{equation}
 Eq.(\ref{ecugg}) for $A,B$ reads
\begin{eqnarray}
&&\lambda^2\frac{d^2A}{d\lambda^2}-2\theta_1\lambda^2
\left(\lambda+\frac{p_0q_0}{2}\right)\frac{dA}{d\lambda}+
\frac{\theta_1}{2}p_0q_0\left(1+\frac{\theta_1}{2}p_0q_0\right)A=0,\label{Ag}\\
&& \lambda^2\frac{d^2A}{d\lambda^2}+2\theta_1\lambda^2
\left(\lambda+\frac{p_0q_0}{2}\right)\frac{dA}{d\lambda}-
\frac{\theta_1}{2}p_0q_0\left(1-\frac{\theta_1}{2}p_0q_0\right)A=0,\label{Bg}
\end{eqnarray}
and for $X^+,X^-$ gives
\begin{eqnarray}
&&X^+=-\frac{1}{\theta_1q_0}\left(\frac{dA}{d\lambda}-\frac{\theta_1}{2}p_0q_0\lambda^{-1}A\right),\label{x+g}\\
&&X^-=-\frac{1}{\theta_1p_0}\left(\frac{dB}{d\lambda}+\frac{\theta_1}{2}p_0q_0\lambda^{-1}B\right).\label{x-g}
\end{eqnarray}
Equations (\ref{Ag},\ref{Bg}) can be  transformed into confluent hipergeometric
equations.
Recall that the Tricomi--Kummer's confluent hipergeometric function
$U(a,c,z)$,  \cite{mos}, is a solution of
\[
z\frac{d^2U}{dz^2}+(c-z)\frac{dU}{dz}-aU=0
\]
and has the asymptotic expansion \cite{mos}
\[
U(a,c,z)\sim z^{-a}\sum_{n\geq 0}(-1)^n\frac{(a)_n(a+1-c)_n}{n!}z^{-n},\,\;
z\rightarrow\infty,\, -\frac{3}{2}\pi<\text{arg}z<\frac{3}{2}\pi,
\]
where
$(\alpha)_n=\Gamma(\alpha+n)/\Gamma(n)$.
One can show that
 \[
A(\lambda)=(\theta_1\lambda^2)^{\mu}U\left(\frac{\mu}{2},\frac{1}{2},\theta_1\lambda^2\right)
\]
where
\[
\mu:=\frac{\theta_1}{2}p_0q_0.
\]

Thus,
\[
A(\lambda)\sim\sum_{n\geq
0}(-1)^n\frac{(\frac{\mu}{2})_n(\frac{\mu+1}{2})_n}{n!}(\theta_1\lambda^2)^{-n},\,\;
\lambda\rightarrow\infty.
\]
For $B$ one only needs to replace in the expression for $A$
the parameters $\theta_1\mapsto -\theta_1$ and $\mu \mapsto -\mu$. Hence
\[
B(\lambda)\sim\sum_{n\geq
0}\frac{(-\frac{\mu}{2})_n(\frac{-\mu+1}{2})_n}{n!}(\theta_1\lambda^2)^{-n},\,\;
\lambda\rightarrow\infty.
\]
{}From (\ref{x+g},\ref{x-g}) one gets the corresponding asymptotic expansions
for $X^+,X^-$.
In terms of the Weber's parabolic  cylinder functions \cite{mos} one has for
example
\[
A(\lambda)=2^{-\frac{\mu}{2}}\left(\sqrt{2\theta_1}\lambda\right)^{2\mu}
\exp\left(\frac{\theta_1}{2}\lambda^2\right)D_{-\mu}\left(\sqrt{2\theta_1}\lambda\right),
\]
  and an analogous expression for $B$ is obtained once $\theta_1$ and
$\mu$ are multiplied by $-1$. Notice the appearence of the Hermite polynomials
$H_n$ and the error function Erf, \cite{mos}, when $\mu\in\Bbb{Z}$.
For example when $\mu+1=-N$ with $N\in\Bbb{N}$ we have
\[
A(\lambda)=\left(\sqrt{2\theta_1}\lambda\right)^{-2(N+1)}H_{N+1}\left(\sqrt{\theta_1}\lambda\right),
\]
so that the first row of $g$ is a polynomial, but the second is not as we
already observed. For example, we have
\[
B(\lambda)=K_N\exp(-\theta_1\lambda^2)\frac{d^N}{d\lambda^N}
\left(\exp(\theta_1\lambda^2)\text{Erfc}\left(\sqrt{-\theta_1}\lambda\right)\right)
\]
where $K_N$ is some normalization constant and $\text{Erfc}=1-\text{Erf}$ is
the complement to the error function,
\[
\text{Erf}(z)=\frac{2}{\sqrt{\pi}}\int_0^zdt \exp(-t^2)
\]
 which is not a polynomial. As we have remarked before, the Galilean
self--similar solutions are always associated to subspaces in the Sato
Grassmannian which never belongs to the Segal--Wilson Grassmannian.

For the $\text{NLS}^{\pm}$ reduction we need $q=\mp p^{\ast}$,  therefore
 \[
{\cal J}^{\pm}g(\lambda^{\ast})^{\dag}{\cal J}^{\pm}=g(\lambda)^{-1}
\]
where ${\cal J}^+=\text{id}$ and ${\cal J}^-=H$.
 Taking into account the Eqs.(\ref{Ag},\ref{Bg},\ref{x+g},\ref{x-g}) this is
fulfilled when $\theta_n=i\tilde\theta_n,\tilde\theta_n\in {\Bbb R}$, the
initial condition $q_0=\mp p^{\ast}$ and
$A(\lambda^{\ast})^{\ast}=B(\lambda)$. Therefore, $\mu=\mp
i\tilde\theta_1/2\,|p_0|^2\in\Bbb{R}$.

Now, we analyse the Galilean invariant solutions of the AKNS equation,
thus we suppose $t_n=0$ for $n>2$, and $b=\theta_n=0$ for $n\geq 0$. This
corresponds to a Galilean self--similar solution of the AKNS hierarchy
evaluated at $t_n=0$ for $n>2$. It will turn out to be singular in $t_2=0$. So
we need to shift $t_2$ in order to avoid it.

The string equation is
\[
\begin{cases}
t_1\partial_0p+2t_2\partial_1p=0\\
t_1\partial_0q+2t_2\partial_1q=0.
\end{cases}
\]
Now
\[
 p(t_0,t_1,t_2)=\exp(2t_0)\tilde p(t_1,t_2), \,\;
q(t_0,t_1,t_2)=\exp(-2t_0)\tilde q(t_1,t_2),
\]
with
\[
 \begin{cases}
2t_1\tilde p+2t_2\partial_1\tilde p=0\\
-2t_1\tilde q+2t_2\partial_1\tilde q=0.
\end{cases}
\]
The solutions to these equations are
\[
\tilde p(t_1,t_2)=\exp\left(-\frac{t_1^2}{2t_2}\right)\hat p(t_2),\,\;
\tilde q(t_1,t_2)=\exp\left(\frac{t_1^2}{2t_2}\right)\hat q(t_2)
\]
where the functions $\hat p, \hat q$ must be fixed in order to have solutions
to the AKNS equation, thus
\[
\begin{cases}
2\partial_2\hat p=-\frac{1}{t_2}\hat p-2\hat p^2\hat q\\
2\partial_2\hat q=-\frac{1}{t_2}\hat q+2\hat p\hat q^2.
\end{cases}
\]
Finally, one finds
\[
\tilde p(t_1,t_2)=\exp\left(-\frac{t_1^2}{2t_2}\right)a t_2^{-ab-1/2},
\,\;
\tilde q(t_1,t_2)=\exp\left(\frac{t_1^2}{2t_2}\right)b t_2^{ab-1/2}.
\]
This is a two parameter family of Galilean self--similar solutions to the AKNS
hierarchy. In fact, when one performs the shift $t_2\rightarrow t_2+\theta_1/2$
one finds  $\mu=ab$, that together with $p_0$ (or $q_0$) parametrizes the
solution. One can see that $\mu$ is the unique non trivial parameter recalling
that if $(p,q)$ is a solution of the AKNS then so is any $(e^cp,e^{-c}q)$.  We
have for the specific heat
\[
\partial_1^2\ln Z(t_1,t_2,0,\dots)=-\frac{\mu}{t_2}.
\]
This is the solution corresponding to the point in the Grassmannian we have
found above.

The corresponding  Galilean self--similar solution to the $\text{NLS}^{\pm}$ is
 of the   form
\[
\tilde p(t_1,t_2)=\exp\left(i\frac{t_1^2}{2t_2}\right)\hat p(t_2),
\]
where $\hat p$ satisfies
\[
2\partial_2\hat p=-\frac{1}{t_2}\hat p\mp 2i|\hat p|^2\hat p.
\]
Writing $\hat p=|\hat p|\exp(i\text{arg}\hat p)$ one obtains the equations
\begin{eqnarray*}
&&\partial_2|\hat p|=-\frac{1}{2t_2}|\hat p|,\\
&&\partial_2\text{arg}\hat p=\mp |\hat p|^2.
\end{eqnarray*}
Therefore
\[
\tilde p(t_1,t_2)=e^{ia}\sqrt{\left|\frac{\mu}{t_2}\right|}
\exp\left(i\left(\frac{t_1^2}{2t_2} \mp|\mu|\text{sgn}t_2\ln
|t_2|\right)\right).
\]
Here $a\in\Bbb{R}$ is an arbitrary phase that can be removed. We have a
1--parameter family of Galilean self--similar solutions to the
$\text{NLS}^{\pm}$ defined for $t_2\neq 0$, with
\[
|p|=\sqrt{\left|\frac{\mu}{t_2}\right|},
\]
so that it vanishes at $t_2\rightarrow\pm\infty$ and generates a singular
behaviour at $t_2=0$, an explode--decay phenomena for a non--localized wave.

\subsection{Scaling self--similarity}
We are going now to consider the string equation corresponding to the vector
field $X=\boldsymbol{\varsigma}+\theta_0\partial_0+\theta_1\partial_1$.
As we have already discussed this corresponds  to
self--similar solutions under a $(1+2\theta_0,1-2\theta_0)$ weighted scaling in
the shifted coordinates $t_1\mapsto t_1+\theta_1$ and $t_n\mapsto t_n$ for
$n>1$. This last shift allows us to avoid possible singularities of the
solution at $t_1=0$.

Let
\[
g=\text{id}+\lambda^{-1}X_1+\cdots
\]
be the initial condition for the commuting flows $\psi({\bf t})$. In order to
have self--similar solutions under the vector field $X$, it must satisfy
\begin{equation}
\lambda\frac{dg}{d\lambda}+g (\theta_1 p_0
E+(\theta_0+\theta_1\lambda)H+\theta_1q_0F)  =
(\theta_0+\theta_1\lambda)Hg\label{ecug}
\end{equation}
which implies for the matrix coefficients $X_n$ of the Laurent expansion of $g$
\[
-nX_n-\theta_0[H,X_n]+\theta_1 X_n(p_0E+q_0F)=\theta_1[H,X_{n+1}].
\]
 With the use of (\ref{not1}) one finds the recurrence laws
\begin{eqnarray*}
&&X_{n+1}^+=\frac{1}{2\theta_1}\left(-n-2\theta_0+\frac{\theta_1^2p_0q_0}{n}\right)X_n^+\\
&&X_{n+1}^-=-\frac{1}{2\theta_1}\left(-n+2\theta_0+\frac{\theta_1^2p_0q_0}{n}\right)X_n^-
\end{eqnarray*}
and
\begin{eqnarray*}
&&A_n=\frac{\theta_1q_0}{n}X_n^+\\
&&B_n=\frac{\theta_1p_0}{n}X_n^-
\end{eqnarray*}
that together with   $X_1^+=p_0/2$ and $X_1^-=-q_0/2$ give us the matrix $g$.
There are cases for which this expansion is a polynomial in $\lambda^{-1}$ and
represents therefore not only an asymptotic expansion but also a well defined
function.  We require
\begin{equation}
\theta_1^2p_0q_0=(N^++2\theta_0)N^+=(N^--2\theta_0)N^-\label{rat}
\end{equation}
with $N^{\pm}\in\Bbb{N}\cup \{ \text{0} \}$, so that
\[
X_n^+,A_n=0,\,\; n>N^+
\]
and
\[
X_n^-,B_n=0,\,\; n>N^-.
\]
Hence, we get a  polynomial $g$ in $\lambda^{-1}$ of degree $N^+$ in the first
row and degree $N^-$ in the second one.
 Eqs.(\ref{rat}) imply
\begin{eqnarray*}
&&2\theta_0=N^--N^+\in\Bbb{Z}\\
&&\theta_1^2p_0q_0=N^+N^-\in\Bbb{N}\cup\{ \text{0} \}.
\end{eqnarray*}
This gives points in Segal--Wilson Grassmannian associated with solutions of
the AKNS hierarchy $(p,q)$ which are self--similar under the
$(1+N^+-N^-,1-N^++N^-)$
weighted scaling symmetry.

Using (\ref{not2}),
  Eq.(\ref{ecug})  for $A,B$ reads
\begin{eqnarray}
&&\lambda^2\frac{d^2A}{d\lambda^2}+((1-2\theta_0)\lambda-2\theta_1\lambda^2)\frac{dA}{d\lambda}-\theta_1^2p_0q_0A=0,\label{A}\\
&&\lambda^2\frac{d^2B}{d\lambda^2}+((1+2\theta_0)\lambda+2\theta_1\lambda^2)\frac{dB}{d\lambda}-\theta_1^2p_0q_0B=0,\label{B}
\end{eqnarray}
and for $X^+,X^-$ we obtain the expressions
\begin{eqnarray}
&&X^+=-\frac{\lambda}{\theta_1q_0}\frac{dA}{d\lambda},\label{x+}\\
&&X^-=-\frac{\lambda}{\theta_1p_0}\frac{dB}{d\lambda}.\label{x-}
\end{eqnarray}
Equations (\ref{A},\ref{B}) are equivalent to confluent hipergeometric
equations.
Consider the roots $(\mu_+,\mu_-)$ of
\[
\mu^2-2\theta_0\mu-\theta_1^2p_0q_0=0,
\]
we get for $\theta_0$ the value
 \[
2\theta_0=\mu_++\mu_-,\,\;\mu_+\mu_-=-\theta_1^2p_0q_0.
\]

If we define
\[
A(\lambda)=\lambda^{\mu_+}U(2\theta_1\lambda),
\]
then $U(z)$ satisfies
\[
z\frac{d^2U}{dz^2}+ (1+\mu_+-\mu_--z)\frac{dU}{dz}-
\mu_+ U=0,
\]
thus we are dealing with  the Tricomi--Kummer's confluent hipergeometric
function  $U(a,c,z)$   with $a=\mu_+$ and $c=1+\mu_+-\mu_-$, and we deduce for
$A(\lambda)$ the behaviour
\[
A(\lambda)\sim\sum_{n\geq
0}(-1)^n\frac{(\mu_+)_n(\mu_-)_n}{n!}(2\theta_1\lambda)^{-n},\,\;
\lambda\rightarrow\infty.
\]

For $B$ the analysis is the same, we only need to replace $2\theta_0$ and
$2\theta_1$ by $-2\theta_0$ and $-2\theta_1$ respectively in the  formulas
above. So the asymptotic expansion for $B$ is
 \[
B(\lambda)\sim\sum_{n\geq 0}
\frac{(-\mu_+)_n(-\mu_-)_n}{n!}(2\theta_1\lambda)^{-n},\,\;
\lambda\rightarrow\infty.
\]
{}From   formulas (\ref{x+},\ref{x-}) we obtain the asymptotic expansions for
$X^+$ and $X^-$. We have
\begin{eqnarray*}
&&X^+(\lambda)\sim\frac{1}{\theta_1q_0}\sum_{n\geq
1}(-1)^n\frac{(\mu_+)_n(\mu_-)_n}{(n-1)!}(2\theta_1\lambda)^{-n},\,\;
\lambda\rightarrow\infty\\
&&X^-(\lambda)\sim\frac{1}{\theta_1p_0}\sum_{n\geq 1}
\frac{(-\mu_+)_n(-\mu_-)_n}{(n-1)!}(2\theta_1\lambda)^{-n},\,\;
\lambda\rightarrow\infty.
\end{eqnarray*}

Let us  notice that  when $\mu_++\mu_-=0$ the  function $U$ can be expressed in
terms of the Macdonalds--Basset function \cite{mos}, for example if
$z=2\theta_1\lambda$ we have
\[
A(\lambda)=\left(1+\mu_+-\frac{d}{dz}\right)\sqrt{\frac{z}{\pi}}
\exp(z/2)K_{\mu_+-1/2}(z/2).
\]

For the $\text{NLS}^{\pm}$ reduction we need  that $\mu_+,\mu_-$   be solutions
of
\[
\mu^2-2i\tilde\theta_0\pm|\tilde\theta_1p_0|^2=0.
\]

In the polynomial case of the AKNS hierarchy we must have (or the other way
around)
\[
\mu_+=-N^+,\,\;\mu_-=N^-.
\]
Again, from the asymptotic expansions, we see that $A,X^+$ and $B,X^-$ are a
polynomials in $\lambda^{-1}$ of degree $N^+$ and $N^-$ respectively.
The solutions in the polynomial case are the rational solutions of the AKNS
hierarchy appearing in \cite{s}. To connect with the notation of that paper we
notice that $1+\text{p}-\text{q}= N^+-N^-$ and that $\text{p}=N^+N^-$ where p,q
are the degree of the polynomials corresponding to the tau functions
$\sigma,\tau$ for the AKNS hierarchy defined in that paper.
 This implies that $\text{q}=(N^+-1)(N^-+1)$, and so $n-k=N^+$ and $k+1=N^-$ or
viceversa  ($n+1=N^++N^-$), where $n,k$ are those of \cite{s}.

One can easily see that the polynomial case described above is the only case
for which the asymptotic series converges and defines a function in a
neighbourhood of $\lambda=\infty$. Therefore, they are the only points in the
Segal--Wilson Grassmannian corresponding to weighted scaling self--similar
solutions,  generically we have points in the Sato Grassmannian.
 Observe that for the  $\text{NLS}^{\pm}$ hierarchies one arrives to the
condition $2\theta_0=N^--N^+$ with $\theta_0\in i\Bbb{R}$, so $\theta_0=0$.
Then $\mu_{\pm}=\pm|\tilde\theta_1p_0|$ in the $\text{NLS}^+$ case and
 $\mu_{\pm}=\pm i|\tilde\theta_1p_0|$ for the $\text{NLS}^-$ case. So that none
of the Sachs rational solutions for the AKNS system reduces to  the
$\text{NLS}^-$ equation, furthermore it is known that this equation does not
have rational solutions. Only for the $\text{NLS}^+$ hierarchy we have points
in the Segal--Wilson Grassmannian corresponding to the reduced Sachs solutions,
the Nakamura--Hirota rational solutions for $\text{NLS}^+$ equation, \cite{nh}.
Now, $N^+=N^-$ and $n=2k+1$.   Notice that in   \cite{nh} it is considered not
only $n=2k$, when they analyse the Boussinesq system,  as was claimed in
\cite{s} but also $n=2k+1$, when they study the $\text{NLS}^+$ equation.

Summing, for the Segal--Wilson case we have

\begin{pro}
The $(n,k)$ rational solution for the AKNS hierarchy found in \cite{s}
corresponds to the point in the Segal--Wilson Grassmannian associated to the
coset $g\cdot L^+SL(2,\Bbb{C})$ where $g\in L^-_1SL(2,\Bbb{C})$ is given by the
following Laurent polynomial
\[
g(\lambda/2\theta_1)=\left(\begin{array}{cc}
\sum_{n= 0}^{N^+}\frac{(-N^+)_n(N^-)_n}{n!}(-\lambda)^{-n}&
  \frac{1}{q_0}\sum_{n= 1}^{N^+}
\frac{(-N^+)_n(N^-)_n}{(n-1)!}(-\lambda)^{-n}\\
 \frac{1}{p_0}\sum_{n= 1}^{N^-}
\frac{(N^+)_n(-N^-)_n}{(n-1)!}(\lambda)^{-n}&
\sum_{n=0}^{N^-}\frac{(N^+)_n(-N^-)_n}{n!}(\lambda)^{-n}
\end{array}\right),
\]
where $n+1=N^++N^-$ and $k+1=N^-$. These are the only weighted scaling
self--similar solutions with a corresponding point in the Segal--Wilson
Grassmannian.
None of these reduce to the $\text{NLS}^-$ hierarchy and only when $N^+=N^-$,
($n=2k+1$), $-p_0^{\ast}=q_0$  they reduce to solutions of the $\text{NLS}^+$
hierarchy.
\end{pro}
\section{Acknowledgements}
One of the authors (MM) is indebted to Dr.P.Guha for initial collaboration and
to Prof.L.Bonora and Prof.G.Wilson for providing their papers.


\newpage
\begin{thebibliography}{99}
\bibitem{ac} M.Ablowitz and P.Clarkson, {\em Solitons, Nonlinear
Evolution Equations and Inverse Scattering} London Math.Soc.Lec.Not.Ser.
{\bf 149}, Cambridge University Press (1991), Cambridge.
\bibitem{akns} M.Ablowitz, D.Kaup, A.Newell, and H.Segur, Phys.Rev.Lett. {\bf
31} (1973) 125; V.Zakharov and A.Shabat, Func.Anal.Appl. {\bf 8} (1974) 43,
Func.Annal.Appl. {\bf 13} (1979) 166.
\bibitem{bk} M.Bergveld and A. ten Kroode, J.Math.Phys. {\bf 29} (1988) 1308.
 \bibitem{her} E.Brezin and V.Kazakov, Phys.Lett. {\bf 236B} (1990) 144;
M.Douglas and M.Shenker, Nucl.Phys. {\bf B335} (1990) 685;  D.Gross and
A.Migdal, Phys.Rev.Lett. {\bf 64} (1990)
127.
\bibitem{bx} L.Bonora and C.Xiong,
Phys.Lett. {\bf 285B} (1992) 191; {\em Matrix models without scaling limit}
Int.J.Mod.Phys.A (1993) to appear.
\bibitem{sta}S.Dalley, C.Johnson, and T.Morris, Nuc.Phys.  {\bf  B368}
(1992) 625, 655;   S.Dalley, preprint PUPT, {\bf 1290} (1991);  C.Johnson,
T.Morris, and A.W\"atterstam, Phys.Lett. {\bf 291B} (1992) 11;
  C.Johnson, T.Morris, and P.White, Phys.Lett. {\bf 292B} (1992) 283; S.Dalley,
C.Johnson, T.Morris, and A.W\"attersman, Mod.Phys.Lett. {\bf A29} (1992) 2753;
A.W\"attersman, Phys.Lett. {\bf 263B} (1991) 51.
\bibitem{d} L.Dickey, {\em Another Example of a $\tau$--Function} in
{\em Hamiltonian Systems, Transformations Groups and Spectral Transform
Methods} edited by
 J.Harnad and J.Marsden, Les publications CMR, Universit\'e de Montr\'eal,
(1990) Montr\'eal; J.Math.Phys. {\bf 32} (1991) 2996.
\bibitem{d2} L.Dickey, Commun.Math.Phys. {\bf 82} (1981) 345.
\bibitem{ds} V.Drinfel'd and V.Sokolov,  J.Sov.Math. {\bf 30} (1985) 1975.
\bibitem{du} B.Dubrovin, Fun.Anal.Appl. {\bf 11} (1977) 265.
\bibitem{ft} L.Faddeev and L.Takhtajan, {\em Hamiltonian Methods in the Theory
of Solitons} Springer Verlag (1987), Berlin.
\bibitem{fnr} H.Flaschka, A.Newell, and T.Ratiu, Physica {\bf 9D} (1983) 300.
\bibitem{gm1} F.Guil and M.Ma\~nas, Lett.Math.Phys. {\bf 19} (1990) 89;
M.Ma\~nas, {\em
Problemas de factorizaci\'on y sistemas integrables} PhD thesis, Universidad
Complutense de Madrid (1991), Madrid.
\bibitem{gm2} F.Guil and M.Ma\~nas, {\em Self--similarity
in the KdV hierarchy. Geometry of
the string equations} in {\em Nonlinear Evolution Equations and Dynamical
Systems'92} edited by V.Mahankov, O.Pashaev, and I.Puzynin, World Scientific
(1992), Singapure;
  {\em Strings equations for the KdV hierarchy and the Grassmannian} J.Phys.A:
Math. \& Gen. (1993) to appear; M.Ma\~nas and P.Guha, {\em String equations for
the unitary matrix model and the periodic flag manifold} (1993) to appear.
\bibitem{ks} V.Kac and A.Schwarz, Phys.Lett. {\bf 257B}
(1991) 329; A.Schwarz, Mod.Phys.Lett. {\bf A6} (1991) 611 and 2713;
K.Anagnostopoulos, M.Bowick, and A.Schwarz, Commun.Math.Phys. {\bf 148} (1992)
148.
\bibitem{mos} W.Magnus, F.Oberhettinger, and R.Soni, {\em Formulas and Theorems
for the Special Functions of Mathematical Physics} Springer Verlag (1966),
Berlin.
\bibitem{nh} A.Nakamura and R.Hirota, J.Phys.Soc.Jpn. {\bf 54} (1985) 491.
\bibitem{n} A.Newell, {\em Solitons in Mathematics and Physics} SIAM (1985),
Philadelphia.
\bibitem{no} S.Novikov, Func.Anal.Appl. {\bf 24} (1991) 296.
\bibitem{un} V.Periwal and D.Shevitz, Phys.Rev.Lett. {\bf 64} (1990) 1326;
Nuc.Phys. {\bf B344} (1990) 731;  K.Anagnostopoulos, M.Bowick, and
N.Ishisbashi, Mod.Phys.Lett. {\bf A6} (1991) 2727.
\bibitem{ps} A.Pressley and G.Segal, {\em Loop groups}
Oxford University Press (1985), Oxford.
\bibitem{pre} E.Previato, Duke Math.J. {\bf 52} (1985) 329.
\bibitem{s} R.Sachs, Physica {\bf D30} (1988) 1;
{\em Polynomial $\tau$--functions for the AKNS hierarchy} in
{\em Theta functions---Bowdoin 1987. Part 1} Proc.Sympos. Pure Maths. {\bf 49},
part 1, 133,  AMS (1989) Providence.
\bibitem{sa} M.Sato, RIMS Kokyuroku {\bf 439} (1981) 30;
{\em The KP hierarchy and infinite--dimensional Grassmann manifolds}  in
{\em Theta functions---Bowdoin 1987. Part 1} Proc.Sympos. Pure Maths. {\bf 49},
part 1, 51,  AMS (1989) Providence.

\bibitem{sw} G.Segal and G.Wilson, Publ.Math.IHES {\bf 61} (1985) 1.
\bibitem{w} G.Wilson, {\em The $\tau$--functions of the g--AKNS hierarchy},
Proceedings of Verdier memorial conference, Y.Kosman--Schwarzbach {\em et al}
Eds. Birkh\"auser (1993) Berlin.
\bibitem{wk} E.Witten, Surv.Diff.Geom. {\bf 1} (1991) 243;
M.Kontsevich, Commun.Math.Phys. {\bf 147} (1992) 1.





















\end{thebibliography}
\end{document}